\documentclass[%
 reprint,
 amsmath,
 amssymb,
 aps,
 nofootinbib
]{revtex4-2}

\usepackage{physics}
\usepackage{graphicx}
\usepackage{dcolumn}
\usepackage{bm}
\usepackage[parfill]{parskip}
\usepackage[usenames,dvipsnames]{xcolor}
\usepackage{hyperref}
\usepackage{cleveref}
\usepackage{nameref}
\hypersetup{
    pdfencoding=unicode,
	colorlinks=true,
	urlcolor=Maroon,
	linkcolor=RoyalBlue,
	citecolor=Maroon,
	pdftitle={},
	pdfauthor={},
	pdfdisplaydoctitle=true,
	pdfstartview=FitH,
	linktocpage=true
}
\usepackage{amsmath}
\usepackage{dsfont}
\usepackage{amsfonts}
\usepackage{amssymb}
\usepackage{mathtools}
\usepackage{microtype}
\usepackage{bm}
\usepackage[utf8]{inputenc}
\usepackage{blkarray}
\usepackage{bigstrut}
\usepackage{nccmath}
\usepackage{float}
\usepackage{braket}
\usepackage{dsfont}
\usepackage[export]{adjustbox}
\usepackage[english]{babel}
\usepackage{stmaryrd}
\usepackage{orcidlink}
\usepackage{tikz}
\usepackage{bbm}
\usepackage{blindtext}
\usepackage{fontawesome}
\usepackage[]{notes2bib}
\usepackage{enumitem}

\usepackage{tikz}
\usetikzlibrary{
  decorations.pathmorphing,
  decorations.pathreplacing,
  decorations.shapes,
  shapes.misc
}
\tikzset{
  cross/.style={
    cross out, draw=black,
    minimum size=2*(#1-\pgflinewidth),
    inner sep=0pt, outer sep=0pt
  },
  cross/.default={1pt}
}
\usetikzlibrary{shapes,arrows,positioning,decorations.markings}
\usetikzlibrary{snakes, patterns,angles,quotes,arrows.meta,shapes.misc}

\usepackage[most]{tcolorbox}
\newtcbox{\mymath}[1][]{
  nobeforeafter, math upper, tcbox raise base,
  enhanced, colframe=blue!30!black,
  boxrule=1pt, #1
}

\DeclareFontFamily{U}{mathx}{}
\DeclareFontShape{U}{mathx}{m}{n}{<-> mathx10}{}
\DeclareSymbolFont{mathx}{U}{mathx}{m}{n}
\DeclareMathAccent{\widehat}{0}{mathx}{"70}
\DeclareMathAccent{\widecheck}{0}{mathx}{"71}
 
\DeclareMathOperator*{\sumint}{%
\mathchoice%
  {\ooalign{$\displaystyle\sum$\cr\hidewidth$\displaystyle\int$\hidewidth\cr}}
  {\ooalign{\raisebox{.14\height}{\scalebox{.7}{$\textstyle\sum$}}\cr\hidewidth$\textstyle\int$\hidewidth\cr}}
  {\ooalign{\raisebox{.2\height}{\scalebox{.6}{$\scriptstyle\sum$}}\cr$\scriptstyle\int$\cr}}
  {\ooalign{\raisebox{.2\height}{\scalebox{.6}{$\scriptstyle\sum$}}\cr$\scriptstyle\int$\cr}}
}

\usepackage[normalem]{ulem}

\usepackage{hyperref}

\usepackage[caption=false]{subfig}

\definecolor{darkblue}{cmyk}{0.9,0.9,0,0}
\definecolor{darkgreen}{cmyk}{0.9,0,0.9,0}
\definecolor{blueblue}{cmyk}{0.73,0.28,0,0.5}
\definecolor{lightblue}{RGB}{55,171,200}
\definecolor{grey}{gray}{0.55}
\definecolor{pink}{cmyk}{0., 0.9859943977591037, 0.3571428571428571, 0.16000000000000003}
\definecolor{lightpink}{cmyk}{0., 0.5, 0.5, 0.}
\definecolor{lightgreen}{cmyk}{0.24175824175824182, 0., 0.9615384615384616, 0.28627450980392155}

\def\be#1\ee{\begin{align}#1\end{align}}

\def\({\left(}\def\){\right)}
\def\[{\left[}\def\]{\right]}
\def\<{\langle}\def\>{\rangle}

\newcommand{\beq}{\begin{equation}}\newcommand{\eeq}{\end{equation}}
\newcommand{\beqq}{\begin{equation*}}\newcommand{\eeqq}{\end{equation*}}
\newcommand\beqa{\begin{eqnarray}}\newcommand\eeqa{\end{eqnarray}}

\begin{document}

\title{Resonances: Universality and Factorization on Higher Sheets}

\author{Miguel Correia}
\author{Celina Pasiecznik}
\affiliation{Department of Physics, McGill University, 3600 Rue University, Montr\'eal, H3A 2T8, QC, Canada}

\begin{abstract}
\noindent  
Most particles in nature are unstable, manifesting as resonances in scattering processes. Using analyticity and unitarity, we show nonperturbatively that resonances, defined as poles on higher Riemann sheets of scattering amplitudes, share basic properties with stable particles: (i) Universality, that a resonance generically appears in every S-matrix element; and (ii) Factorization, that amplitudes factorize on resonance poles. Our framework applies in any spacetime dimension and across arbitrarily many two-particle cuts, including cases where the kinematic Riemann surface becomes infinitely sheeted. Importantly, we find that resonance data (mass, width, couplings, and sheet index) are fully encoded on the physical sheet, where causality can impose additional constraints. These results are relevant for extending S-matrix bootstrap studies beyond elastic scattering.
\end{abstract}

\maketitle

\section{Introduction}

Most particles observed experimentally are unstable \cite{ParticleDataGroup:2024cfk}. They do not belong to the asymptotic Fock space, but rather appear as short-lived enhancements in scattering processes, traditionally modeled by Breit--Wigner--type shapes. Extracting the properties of resonances from experimental data is notoriously delicate \cite{JPAC:2021rxu,Mai:2022eur}: a broad resonance may leave only a mild imprint on experimental data, and conversely not every bump corresponds to a true resonance, since Landau singularities \cite{Landau:1959fi}, which are of purely kinematic origin, can also induce rapid variations of the amplitude \cite{Mikhasenko:2015oxp}. This tension between ``bumps'' and genuine poles makes it natural to ask, in a fully nonperturbative setting, what is the rigorous definition of a resonance and how close such an object is to an actual particle.

From the viewpoint of quantum field theory, the notion of a particle is tightly linked to the analytic structure of the S-matrix \cite{chew1966analytic}. Stable particles correspond to simple poles of the amplitude on the physical sheet. Unitarity, when applied to intermediate one-particle states, implies two familiar consequences: amplitudes factorize on the pole into a product of lower-point amplitudes, and the same one-particle state necessarily appears in every channel to which it can couple. We refer to these two properties as \emph{factorization} and \emph{universality}. They are standard in perturbation theory, where they can be derived diagrammatically from Feynman rules, and they underlie modern on-shell methods to bootstrap tree-level amplitudes in gravity, Yang--Mills, and related theories \cite{Cheung:2017pzi}.

Unstable states modify this picture in an essential way. Their poles move off the real axis and onto unphysical Riemann sheets reached by analytic continuation across unitarity cuts. A resonance is then identified with a pole of the scattering amplitude on such a higher sheet \cite{Gribov:2009cfk}. Singularities on these higher sheets can influence the amplitude on the physical sheet; a pole on the second sheet, for instance, may sit close enough to the cut to generate a rapid variation of the amplitude that is observable in scattering experimental data. This naturally raises the question of whether resonances enjoy the same universality and factorization properties as stable particles.

The nonperturbative S-matrix bootstrap \cite{Paulos:2016but,Paulos:2017fhb,Kruczenski:2022lot} provides a complementary motivation. In practice, implementations work directly with amplitudes on the physical sheet, imposing unitarity along the branch cuts through inequalities and sum rules. Resonances, understood as higher-sheet poles, are then inferred indirectly: in elastic scattering they correspond to zeros of the partial-wave amplitude on the first sheet \cite{Doroud:2018szp,Guerrieri:2018uew}. It is not obvious, however, that this strategy extends beyond the simplest elastic case. To push the S-matrix bootstrap beyond the purely elastic regime, it is therefore crucial to understand whether, and in what sense, resonance data is encoded in the physical sheet. This data consists of the mass, width, couplings, and the sheet on which the resonance resides.

In this paper we rigorously address these questions using analyticity and unitarity. We show that resonances share the universality and factorization properties familiar from stable particles. Schematically, on the pole of a resonance `$\mathrm{R}$' we find
\begin{align}
&T^{(\mathrm{n})}_{ab \to cd}(s\to s_\mathrm{R}) = \raisebox{-0.4\height}{\includegraphics[width=2.1cm]{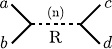}} \;\sim\;
  \frac{g_{ab\mathrm{R}}\, g_{cd\mathrm{R}}}
       {s - s_\mathrm{R}} \,,  
\end{align}
with $s_{\mathrm{R}} \in \mathbb{C}$ in the unphysical sheet  with index $\mathrm{n}$.

Moreover, we derive general relations between the resonance data $\{\mathrm{n},s_\mathrm{R}, g_{a b R}\}$ and the amplitude $T_{ab \to cd}$ on the physical sheet, ready to be used for S-matrix bootstrap applications. Our analysis applies beyond the simplest elastic second-sheet setup, extending to higher sheets connected through any combination of two-particle production thresholds. In particular, it also applies to theories in odd-dimensional spacetimes or involving massless particles where the kinematic Riemann surface can become infinitely sheeted.

Our work is organized as follows. In Sec.~\ref{sec:uniana} we derive a generalized K-matrix solution, which solves analyticity and unitarity across two-particle thresholds. In  Sec.~\ref{sec:anacont} we show how this general solution can be analytically continued to higher Riemann sheets and use it to derive properties of resonance poles and relate them to the physical sheet amplitude. In Sec.~\ref{sec:examples} we showcase our relations in a system of two different particles with $\mathbb{Z}_2$ symmetry.  We conclude in Sec.~\ref{sec:conc} with a brief summary and an outlook on future directions.

\section{Unitarity and Analyticity}
\label{sec:uniana}

\noindent Our starting point is the S-matrix operator $\hat{S}$, which can be split in the usual way in terms of a free and interacting part $\hat{T}$,
\begin{equation}
    \hat{S} = \hat{1} + i \,\hat{T}.
\end{equation}

We will be looking at $2 \to 2$ scattering elements,
\begin{equation}
    \bra{p_a,p_b} \hat{T} \ket{p_c,p_d} = (2\pi)^d \delta^d(p_a+p_b-p_c-p_d)\, T_{ab \to cd}(s,t)
\end{equation}
where $T_{ab \to cd}(s,t)$ is the scattering amplitude and depends on Mandelstam invariants $s = (p_a+p_b)^2$ and $t = (p_a-p_c)^2$.  
Each momenta satisfies the on-shell relation $p_i^2 = m_i^2$, where $m_i$ is the corresponding mass. While we consider the scattering of scalar particles, our analysis generalizes straightforwardly to spinning particles.

\textbf{Partial-wave amplitudes.} We introduce definite angular-momentum $J$ scattering elements, known as partial-wave amplitudes,
\begin{equation}
    \label{eq:pw}
T^J_{ab \to cd}(s) = {\mathcal{N}_d \over 2} \int_{-1}^1 (1-z^2)^{d-4 \over 2}P^{(d)}_J(z) \, T_{ab \to cd}(s,t(z)) \, dz,
\end{equation}
where $P^{(d)}_J(z)$ is the Gegenbauer polynomial (which reduces to the Legendre polynomial for $d=4$), and $z$ is the cosine of the scattering angle in the center-of-mass frame given in terms of $s$ and $t$ (see Appendix \ref{sec:kin}).

The inverse relation reads
\begin{equation}
\label{eq:Tpwd}
    T_{ab\to cd}(s,t) = \sum_{J=0}^\infty n_J^{(d)} \,T^J_{ab \to cd}(s) \, P_J^{(d)}(z)\,.
\end{equation}
The normalization factors are
\begin{equation}
    \mathcal{N}_d = {(16\pi)^{2 -d \over 2} \over \Gamma({d -2 \over 2})}, \;n_J^{(d)}\!= \!\frac{ (4\pi)^{\tfrac{d}{2}} (d + 2J - 3)\, \Gamma(d + J - 3) }
{ \pi\, \Gamma\!\left(\tfrac{d-2}{2}\right)\, \Gamma(J+1) }
\end{equation}
where we adopt conventions from \cite{Correia:2020xtr}.

Particles manifest as simple poles of scattering amplitudes, and  are most cleanly identified in the partial-wave decomposition. Concretely, a particle of mass $M$ and spin $J'$ produces a simple pole at $s = M^2$ in the partial-wave amplitude of spin $J=J'$,\footnote{The remaining partial waves with spin $J' \neq J$ are regular at $s = M^2$. Hence, in the expansion \eqref{eq:Tpwd} the spin-$J'$ partial wave dominates as $s \to M^2$, and the amplitude develops the simple pole
$T_{ab \to cd}(s \to M^2,t(z)) \propto \frac{P_{J}^{(d)}(z)}{s - M^2}$ in agreement with \cite{Arkani-Hamed:2017jhn}.}

\begin{equation}
   T^J_{ab \to cd}(s \to M^2) = - {g_{abM}g_{cdM} \over s - M^2} \delta_{J,J'}
\end{equation}
where $g_{abM}$ are the cubic couplings between particles $a$, $b$ and $M$.

\textbf{Unitarity.} The S-matrix satisfies unitarity $\hat{S}^\dagger \hat{S} = \hat{1}$ which in terms of $\hat{T}$ reads
\begin{equation}
\label{eq:uniop}
    2\,\mathrm{Im} \,\hat{T} =  \hat{T}^\dagger \hat{T}\,.
\end{equation}
Evaluating this relation between \emph{in} and \emph{out} two-particle momentum states, and inserting a complete set of particle states in-between $\hat{T}^\dagger T$ we find the integral form of unitarity. In schematic form,
\begin{equation}
\label{eq:uniT}
   2\, \mathrm{Im}\, T_{ab \to cd} =  \sumint_{ij} T^*_{ab \to ij} T_{ij\to cd} + \sumint_{n\geq 3} T_{\{2 \to n\}}^* T_{\{2 \to n\}}
\end{equation}
where $\sumint$ denotes summing over intermediate particle states and integrating over their phase-space. The first piece corresponds to the contribution of any two-particle states, with the integral running over the angles of intermediate $i,j$ particles (see Appendix \ref{sec:kin}), and the second piece denotes all multi-particle contributions, with $n \geq 3$ the number of intermediate particles.

Unitarity takes a simpler form in terms of the partial waves \cite{Gribov:2009cfk,Correia:2020xtr}. We have
\begin{equation}
\label{eq:uniP}
 \mathrm{Im} \,  T_{ab \to cd}^J =\sum_{i>j} \rho_{ij}\, (T_{ab \to ij}^J)^* T^J_{ij\to cd} + \Delta_{\text{HP},ab\to cd}^J 
\end{equation}
where $\Delta_{\text{HP},ab\to cd}^J$ is the higher-point contribution coming from the second piece of \eqref{eq:uniT}, which contains non-trivial integrals in terms of higher-point partial waves. 
\begin{figure*}[t]
  \centering\resizebox{0.8\textwidth}{!}{\begin{tikzpicture}[font=\small]
    \coordinate (origin) at (0,0);
    
    \draw[thick,black] (8.6,1.6) node[] {$s$};
    \draw[thick,black,-] (8.4,1.35) -- ++ (0,0.35);
    \draw[thick,black,-] (8.4,1.35) -- ++ (0.35,0);
    
    \draw[thick,black,->,>=stealth] (origin) -- ++(9.1,0);
    \draw[thick,black,->,>=stealth] (origin) -- ++(0,1.8);
    \draw[thick,black,-] (origin) -- ++(-3,0);
    \draw[thick,black,-] (origin) -- ++(0,-1.5);
    
    \filldraw[gray!50] (1,0) circle (1.5pt);
    \filldraw[gray!50] (1.5,0) circle (1.5pt);
    \draw[-,snake=snake,thick, gray!50, segment amplitude=.6mm,segment length=3mm,line after snake=1mm]
        (1.5,0) to (-3,0);

    \draw[-,snake=snake,thick, red, segment amplitude=.6mm,segment length=2mm,line after snake=1mm]
        (0,0) to (-3,0);
    \filldraw[gray!50] (0,0) circle (1.5pt);
    \filldraw[gray!50] (0.5,0) circle (1.5pt);
    \filldraw[red] (0.8,0) circle (1.5pt); 
    \filldraw[red] (1.2,0) circle (1.5pt);

    \draw[-,snake=snake,thick, blue, segment amplitude=.6mm,segment length=2mm,line after snake=1mm]
        (2.6,0) to (9,0);
    \filldraw[blue] (2.6,0) circle (1.5pt);
    \node at (2.6,-0.3) {$2$P};
    \filldraw[blue] (3.4,0) circle (1.5pt);
    \filldraw[blue] (4.2,0) circle (1.5pt);
    \filldraw[blue] (5.0,0) circle (1.5pt);
    \filldraw[blue] (5.8,0) circle (1.5pt);

    \draw[-,snake=snake,thick, red, segment amplitude=.6mm,segment length=2mm,line after snake=1mm]
        (6.05,0) to (9,0);
    \filldraw[red] (6.0,0) circle (1.5pt);
    \node at (6.0,-0.3) {HP};

    \node at (-1.9,0.9) {$s=0$};
    \draw[->, >=stealth, bend left] (-1.4,0.9) to (-0.1,0.1);
    \node at (-2.2,1.5) {$(m_i-m_j)^2$};
    \draw[->, >=stealth, bend left] (-1.4,1.4) to (0.4,0.1);

    \node at (-1,-0.3) {LC};
 
    \filldraw[gray!50] (3,1.5) circle (1.5pt);
    \node at (3.4,1.5) {$s_R$};

  \node[align=center] at (1.34,1.2) {physical\\sheet};

    \draw
  (1.8,1) to[bend left=20] (2.2,-0.3);

    \draw
    (2.2,-0.3) arc (180:360:0.4) -- (3,-0.3);
    \draw
    (3,-0.3)--(3,0);
    \draw[->, >=stealth]
    (3,-0.3)--(3,-0.1);
    \draw[dashed]
    (3,0) -- (3,1.45);

    \draw[->, >=stealth]
    (3,1.42) -- (3,1.45);
    
    \draw[
      decorate,
      decoration={brace, amplitude=4pt}
    ] (2.6,0.1) -- (6,0.1)
      node[midway, yshift=10pt] {$(m_i+m_j)^2$};

    \draw[
      decorate,
      decoration={brace, amplitude=4pt}
    ] (1.2,-0.1) -- (0.8,-0.1);
    \node at (1,-0.5) {$1$P};

\end{tikzpicture}}
\caption{Analytic structure of the partial-wave S-matrix. In blue: two-particle branch-cuts (2P) with branch-points at $s=(m_i+m_j)^2$ including elastic and production processes. In red: all remaining singularities, including higher-point branch-cuts (HP), single-particle poles (SP) and the partial-wave left-cut (LC). In gray: singularities that emerge in higher Riemann sheets after crossing two-particle cuts. Namely, Landau singularities at $s = 0$ and $s = (m_i-m_j)^2$ and resonances at $s = s_R \in \mathbb{C}$.}
  \label{fig}
\end{figure*}
Conversely, the first contribution in \eqref{eq:uniT} admits a significant simplification. One is left with a sum over two-particle states $ij$, weighed by a phase-space volume factor, which in $d$ spacetime dimensions is given by (see Appendix \ref{sec:kin})
\begin{equation}
\label{eq:rhod}
\rho_{ij}(s) =  b_{ij} \, {\big(s - (m_i-m_j)^2\big)^{d-3 \over 2} (s - (m_i+m_j)^2\big)^{d-3 \over 2} \over s^{d-2 \over 2} }
\end{equation}
where $b_{ij} = 1-\delta_{ij}/2$ is the Bose symmetry factor, and $\rho_{ij}(s)$ is evaluated for center-of-mass energies above the production threshold, $\sqrt{s} \geq m_i+m_j$.

\textbf{Partial-wave S-matrix.} With the partial-wave amplitude elements $T^J_{ab \to cd}$ we can also define a symmetric partial-wave S-matrix:
\begin{equation}
\label{eq:S}
    S_{ab \to cd}^J \equiv \delta_{ab,cd} +2 i \sqrt{\rho_{ab}} \,T^J_{ab \to cd} \,\sqrt{\rho_{cd}}
\end{equation}
which satisfies the unitarity relation
\begin{equation}
\label{eq:uniS}
    \sum_{i>j} (S_{ab \to ij}^J)^* S_{ij \to cd}^J = \delta_{ab,cd} - 4 \sqrt{\rho_{ab}} \,\Delta_{\text{HP},ab\to cd}^J \sqrt{\rho_{cd}}\,.
\end{equation}

\textbf{Analyticity.} Amplitudes are analytic functions, satisfying Hermitian analyticity \cite{olive1963unitarity},\footnote{Note that this property does not generally hold for the S-matrix $S^J_{ab \to cd}(s)$ due to the imaginary factor $i$ in \eqref{eq:S}.}
\begin{equation}
\label{eq:hermitian}
    [T^J_{ab \to cd}(s)]^* = T^J_{ab \to cd}(s^*)\,.
\end{equation}
According to the Feynman $i\epsilon$ prescription, physical amplitudes at real energies are evaluated slightly above the real axis, $s \to s + i\epsilon$ with $\epsilon > 0$ arbitrarily small. Combined with \eqref{eq:hermitian}, we find
\begin{equation}
    \mathrm{Im}\, T^J_{ab \to cd}(s + i\epsilon)
    = \mathrm{Disc}\, T^J_{ab \to cd}(s)
\end{equation}
where the discontinuity across the real axis is
\begin{equation}
\label{eq:disc}
    \mathrm{Disc}\, T^J_{ab \to cd}(s)
    \equiv
    \frac{T^J_{ab \to cd}(s + i\epsilon)
          - T^J_{ab \to cd}(s - i\epsilon)}{2i}\, .
\end{equation}
This expresses the usual statement that unitarity \eqref{eq:uniP} constrains the discontinuity of the amplitude on the real axis. 
Strictly speaking, unitarity is valid only in the physical scattering region,\footnote{Here we assume $m_a + m_b \geq m_c + m_d$ without loss of generality.} namely for $s \geq (m_a + m_b)^2$. However, we will adopt an \emph{extended} unitarity relation that incorporates all lighter two-particle thresholds $m_i + m_j < m_a + m_b$ appearing in \eqref{eq:uniP}. This property holds in perturbation theory, and we will assume it here.

We will further extend the unitarity relation into the region below all two-particle thresholds, going over  single-particle exchanges (which appear as poles of the amplitude) and continuing to negative $s$ toward the left-hand cut, which is dictated by crossing symmetry \cite{Gribov:2003nw}.

With these extensions in place, we write the following unitarity relation, valid for all real $s$:
\begin{equation}
\label{eq:uniF}
   \mathrm{Disc}\, T^J_{ab \to cd}
    =
    \sum_{ij} \Theta_{ij} \,\rho_{ij} \,
         \bigl[T_{ab \to ij}^J\bigr]^*
         T_{ij \to cd}^J
    + \Delta^J_{ab \to cd}
\end{equation}
where $\Theta_{ij} \equiv \Theta(\sqrt{s} - (m_i+m_j))$ and
\begin{equation}
\label{eq:Delta}
    \Delta^J_{ab \to cd}
    \equiv
    \Delta^J_{\mathrm{LC},\,ab \to cd}
    +
    \Delta^J_{\mathrm{SP},\,ab \to cd}
    +
    \Delta^J_{\mathrm{HP},\,ab \to cd}\,.
\end{equation}
The three pieces of $\Delta^J_{ab \to cd}(s)$ are:
\begin{itemize}
    \item $\Delta^J_{\mathrm{LC},\,ab \to cd}(s)$, the contribution from the left-hand cut, supported for $s \leq s_{\mathrm{LC}} <(m_a+m_b)^2$;
    \item $\Delta^J_{\mathrm{SP},\,ab \to cd}(s) = \sum_i g_{abi} g_{cdi}\, \delta(s - m_i^2)$, the contribution from single-particle poles;
    \item $\Delta^J_{\mathrm{HP},\,ab \to cd}(s)$, the contribution from higher-point intermediate states, already present in the physical unitarity relation \eqref{eq:uniP}.
\end{itemize}

These contributions are illustrated in Figure~\ref{fig}.

Crucially, we will not require the explicit form of $\Delta^J_{ab \to cd}$ and will treat it as an unknown throughout.

\textbf{Generalized $\mathrm{K}$-matrix solution.} We will now solve the extended unitarity relation  \eqref{eq:uniF} for the amplitude $T^J_{ab \to cd}$ in terms of the unknown $\Delta^J_{ab \to cd}$. Our approach is based on the K-matrix formalism \cite{ParticleDataGroup:2024cfk,JPAC:2021rxu,Mai:2022eur}, but crucially we do not assume any particular parametrization of the K-matrix.

We first write equation \eqref{eq:uniF} in matrix form
\begin{equation}
\label{eq:uniF2}
    \mathrm{Disc}\, \mathrm{T}
    = (\mathrm{T})^{\dagger} \!\cdot \rho  \cdot \!\mathrm{T}
    +\mathrm{\Delta}
\end{equation}
where, in terms of indices, we treat $\{ab\} = \{ba\}$ as a single index denoting each two-particle state, so that
\begin{equation}
       [\mathrm{T}]_{ab,cd} \equiv T^J_{ab \to cd},\;\; \text{ and } \;\; [\mathrm{\Delta}]_{ab,cd} \equiv \Delta^J_{ab \to cd}\,.
\end{equation}
where $\Delta$ is a positive semi-definite matrix. Moreover,
\begin{equation}
\label{eq:rhoab}
    [\rho]_{ab,ij} = \delta_{ab,ij} \,\rho_{ij} 
\end{equation}
is a diagonal matrix of two-body phase-space volumes. Crossing symmetry, $\{ab \leftrightarrow cd\}$, implies that both $\mathrm{T}$ and $\mathrm{\Delta}$ are symmetric matrices. 

Note that the discontinuity operation \eqref{eq:disc} obeys similar algebraic properties as the derivative operation. In particular,
\begin{equation}
    \mathrm{Disc} \,\mathrm{T}^{-1} = - \,(\mathrm{T^\dagger})^{-1} \cdot(\mathrm{Disc} \,\mathrm{T}) \cdot \mathrm{T}^{-1}\,.
\end{equation}
Using this relation, unitarity \eqref{eq:uniF} can be recast as
\begin{equation}
\label{eq:uniF3}
    -\mathrm{Disc} \,\mathrm{T}^{-1} = \rho + \tilde{\Delta}, \;\; \text{with} \;\; \tilde{\Delta} =  \, (\mathrm{T^\dagger})^{-1} \! \cdot \Delta \cdot \mathrm{T}^{-1}\,.
\end{equation}
It then follows that
\begin{equation}
    \mathrm{Disc} \,( \mathrm{T}^{-1} + \Sigma) = - \tilde{\Delta}
\end{equation}
where $\Sigma$ is a diagonal matrix $[\Sigma]_{ab,ij} = \delta_{ab,ij} \,\Sigma_{ij}$ where each entry satisfies
\begin{equation}
\label{eq:discsigma}
    \mathrm{Disc} \,\Sigma_{ij}(s) = \Theta_{ij}(s)\, \rho_{ij}(s)
\end{equation}
with $\rho_{ij}(s)$ given by \eqref{eq:rhod} and $\Theta_{ij}(s) = \Theta(\sqrt{s}-m_i-m_j)$.

In other words, $\Sigma_{ab}(s)$ is a function whose imaginary part is given by the two-body phase-space volume. Equivalently, $\Sigma_{ab}(s)$ is thus given by the one-loop bubble  Feynman integral, up to analytic terms. 

Finally, we define the K-matrix via the relation
\begin{equation}
\label{eq:kmatrix}
    \mathrm{K}^{-1} \equiv \mathrm{T}^{-1} + \Sigma,
\end{equation}
whose matrix elements obey, using \eqref{eq:uniF3},
\begin{equation}
\label{eq:discK}
    \mathrm{Disc} [\mathrm{K}^{-1}]^J_{ab \to cd}(s)= - \tilde{\Delta}^J_{ab \to cd}(s).
\end{equation}
From \eqref{eq:Delta} we see that the discontinuity of $\mathrm{K}^{-1}$ receives contributions only from higher–point cuts, single–particle exchanges, and the left–hand cut.

In particular, equation \eqref{eq:discK} guarantees that the K-matrix is analytic across the two-particle branch-cuts. Putting everything together, we arrive at the generalized K-matrix solution for the partial-wave amplitude,
\begin{equation}
\label{eq:Kmatrixsol}
    \mathrm{T}(s) = [\mathrm{K}^{-1}(s) - \Sigma(s)]^{-1}.
\end{equation}
The key property of this relation is that it \emph{separates} the analytic structure of the amplitude $\mathrm{T}(s)$, isolating the contribution of two-particle branch-cuts into $\Sigma(s)$, while all remaining singularities (higher-point and left-cut contributions) are grouped into the K-matrix. See Figure \ref{fig} for a representation.

Solution \eqref{eq:Kmatrixsol} admits a natural quantum–mechanical interpretation:
$\Sigma(s)$ plays the role of the two-body Green’s function (or propagator), while $\mathrm{K}(s)$ acts as an effective potential. See \cite{Correia:2024jgr,Oller:2025leg} for a related discussion.

The expression \eqref{eq:Kmatrixsol} represents the most general solution to two-particle unitarity, with all remaining contributions encoded implicitly in the K-matrix. In nuclear phenomenology, it is common to impose strong assumptions on $\mathrm{K}^{-1}(s)$ when fitting experimental data, for example by neglecting all additional branch cuts and treating $\mathrm{K}^{-1}(s)$ as a matrix of polynomial or rational functions. This procedure is used to extract resonances from data and becomes quite delicate for broad resonances (see \cite{JPAC:2021rxu,Mai:2022eur} for recent reviews).\footnote{The K-matrix solution also plays a key role in lattice QCD, namely in translating finite-volume energy spectrum into scattering data \cite{Luscher:1990ux,Briceno:2017max}.}

For our purpose, which requires analytic continuation into higher Riemann sheets across two-particle thresholds, it is enough to note that $\mathrm{K}^{-1}(s)$ is analytic at those thresholds. Consequently, when we cross any two-particle branch cut the function $\mathrm{K}^{-1}(s)$ remains unchanged. This will allow us to access higher Riemann sheets without needing to assume any particular form for the K-matrix.

We conclude this section with a few remarks. Appendix~\ref{sec:cdd} discusses the relation between the generalized K-matrix solution \eqref{eq:Kmatrixsol} and the Castillejo–Dalitz–Dyson (CDD) \cite{Castillejo:1955ed} solution of elastic unitarity, familiar from integrable theories in $d=2$. In this appendix we also show how the K-matrix solution \eqref{eq:Kmatrixsol} can be generalized to incorporate form factors and two-point correlation functions, where Watson’s theorem \cite{Watson:1952ji} immediately follows. 

Note that we have not discussed here the possibility of anomalous thresholds, which can appear on the physical sheet of the scattering amplitude and are not manifestly captured by unitarity \eqref{eq:uniP}. However, as discussed in \cite{Correia:2022dcu}, in the partial-wave amplitude the anomalous threshold is already present in the physical sheet, belonging to the left cut of the K-matrix (depicted in red in Figure \ref{fig}), and is thus taken into account automatically. See also \cite{Aoki:2023tmu} for recent work in this context.

\section{Higher sheets and resonances}
\label{sec:anacont}

As previously mentioned, the generalized K-matrix solution \eqref{eq:Kmatrixsol} manifests the analytic structure across two-particle branch-cuts explicitly via $\Sigma(s)$ whose discontinuity is fixed by the two-particle phase-space volume, according to \eqref{eq:discsigma}. 

This allows us to perform the analytic continuation from the first (physical) sheet through any combination of two-particle cuts into higher (unphysical) sheets. Importantly, $\mathrm{K}^{-1}(s)$ remains unchanged upon this analytic continuation procedure, as long as no other branch-cuts, apart from two-particle cuts, are crossed.

\textbf{Analytic continuation to higher sheets.} On a sheet `$\mathrm{n}$', reached by crossing an arbitrary combination of two–particle cuts, we write
\begin{equation}
\label{eq:Kmatrixsoln}
    \mathrm{T}_{(\mathrm{n})}(s) = [\mathrm{K}^{-1}(s) - \Sigma_{(\mathrm{n})}(s)]^{-1}.
\end{equation}
Here $\mathrm{n}$, which we call the \emph{sheet index}, is a diagonal matrix where each entry $n_{ij}$ records how many times each $ij$-branch-cut has been crossed: a positive (negative) value corresponds to a clockwise (counterclockwise) excursion around the $ij$ branch-cut. 

The matrix $\Sigma_{(\mathrm{n})}(s)$ is also diagonal, with entries
\begin{equation}
    [\Sigma_{(\mathrm{n})}]_{ab,ij} = \delta_{ab,ij} \,\Sigma^{(n_{ij})}_{ij}
\end{equation}
where $\Sigma^{(n_{ij})}_{ij}(s)$ is obtained by analytically continuing $\Sigma_{ij}(s)$ to the sheet labeled by $n_{ij}$, with $n_{ij} = 0$ denoting the physical sheet.

This analytic continuation is precisely where the contrast between massive and massless exchanges, as well as between even and odd spacetime dimensions, becomes sharp: depending on the case, the branch point at $s = (m_i + m_j)^2$ is of square-root or logarithmic type. Inspection of \eqref{eq:rhod} shows that
\begin{itemize}
    \item \textbf{Square-root type:} Occurs for $d$ even with two massive particles, $m_i \neq 0$ and $m_j \neq 0$, or for $d$ odd with two massless particles, $m_i = m_j = 0$. In this case the associated Riemann surface has only two sheets and $n_{ij} \in \{0,1\}$.
    \item \textbf{Logarithmic type:} Occurs for $d$ odd with two massive particles, $m_i \neq 0$ and $m_j \neq 0$; for $d$ even with two massless particles, $m_i = m_j = 0$; and, in any $d$, when exactly one of the particles is massless, e.g.\ $m_i = 0$ and $m_j \neq 0$. In this case the associated Riemann surface is infinitely sheeted, with $n_{ij} \in \mathbb{Z}$.
\end{itemize}

In practice, we will need the monodromy between a higher sheet and the physical sheet, given by the difference $\Sigma^{(n_{ij})}_{ij}(s) - \Sigma_{ij}(s)$. The monodromy between two adjacent sheets follows from analytically continuing the relation~\eqref{eq:discsigma} into the complex plane. Starting from the upper half-plane and encircling the $s = (m_i+m_j)^2$ branch-point counter-clockwise we see that $\Sigma(s - i \epsilon)$ enters the branch cut and moves to sheet $(+1)$, while $\Sigma(s + i \epsilon)$ remains on the physical sheet $(0)$. We thus obtain
\begin{equation}
\label{eq:mon1}
    \Sigma^{(1)}_{ij}(s) - \Sigma_{ij}(s) = -2 i\, \rho_{ij}(s), \;\;\; \mathrm{Im}\, s >0,
\end{equation}
with $\Sigma_{ij} = \Sigma^{(0)}_{ij}$, where $\mathrm{Im}\, s >0$ specifies the branch of $\rho_{ij}(s)$ in \eqref{eq:rhod}, which may itself have a branch-cut starting at $s = (m_i + m_j)^2$.

Indeed, this occurs when the branch-point of $\Sigma_{ij}(s)$ is square-root type (see above), and the monodromy operation flips its sign, $\rho_{ij}(s) \to - \rho_{ij}(s)$. This implies that a second monodromy cycle of \eqref{eq:mon1} returns us to the second-sheet, $\Sigma^{(2)}_{ij}(s) = \Sigma_{ij}(s)$, confirming that only two sheets exist in this case.

\begin{figure}[t]
    \centering
\includegraphics[scale=0.55]{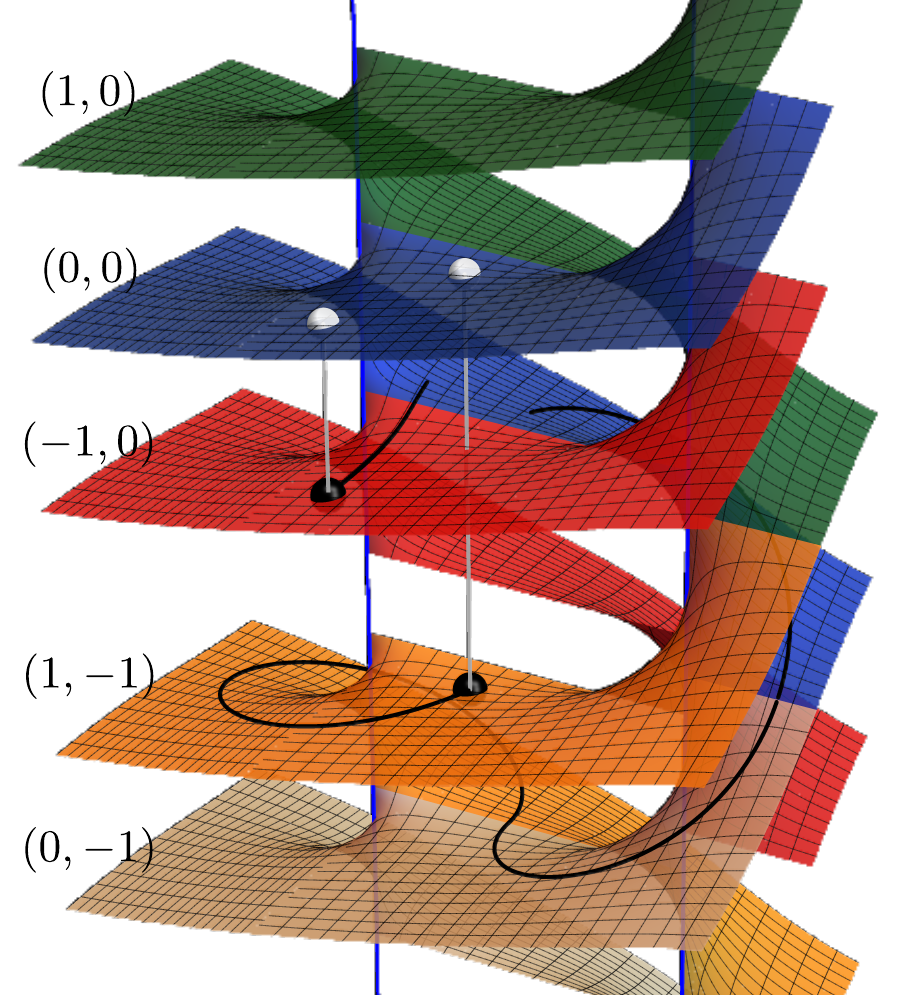}
    \caption{Multi-sheeted Riemann surface generated by a pair of two-particle thresholds with logarithmic branch points $s_{11} < s_{22}$ represented by the vertical blue lines. Sheets are labeled by $(n_{11},n_{22})$ with each number $n_{ij}\in\mathbb{Z}$ indicating the sheet of the corresponding branch-cut, according to \eqref{eq:mon2}. The colored surfaces depict several sheets of the infinite Riemann surface. Black curves show explicit analytic-continuation paths that cross cuts, wind around branch points and thus move between sheets.
The black markers indicate resonances on unphysical higher-sheets with the corresponding translucent markers indicating where their information is encoded on the physical sheet (blue) according to formulas \eqref{eq:mastS1} and \eqref{eq:mastS2}.
}
    \label{fig2}
\end{figure}

By contrast, when the branch-point of $\Sigma_{ij}(s)$ is of logarithmic type, we see that the phase-space factor itself $\rho_{ij}(s)$ in \eqref{eq:rhod} has no branch-point and we can analytically continue \eqref{eq:mon1} further without any obstruction. After $n_{ij}$ counter-clockwise cycles we find the monodromy between sheet $n_{ij}$ and the first sheet
\begin{equation}
\label{eq:mon2}
        \Sigma^{(n_{ij})}_{ij}(s) - \Sigma_{ij}(s) = - 2 i \,n_{ij} \, \rho_{ij}(s)\,,
\end{equation}
whereas for $|n_{ij}|$ clockwise cycles we take $n_{ij} < 0$ in the above.

In summary, the monodromy relation \eqref{eq:mon2} uniformly describes both square-root and logarithmic behaviors. For logarithmic branch points one has $n_{ij} \in \mathbb{Z}$, while for square-root type $n_{ij} \in \{0,1\}$. In the latter case, since $\rho_{ij}(s)$ in \eqref{eq:rhod} has a branch cut for $s \leq (m_i + m_j)^2$, we must also specify which half-plane $s$ lies in: \eqref{eq:mon1} holds for $\operatorname{Im}s > 0$, whereas for $\operatorname{Im}s < 0$ one has $\rho_{ij}(s) \to -\rho_{ij}(s)$.

Repeating this procedure for each $ij$ two-particle cut, and denoting by $(\mathrm{n})$ the full set of integers $n_{ij}$, specifying which branch-cuts have been crossed and how many times, we obtain the desired analytic continuation $\Sigma_{(\mathrm{n})}(s)$ which we write in the compact form
\begin{equation}
\label{eq:monf}
    \Sigma_{(\mathrm{n})} - \Sigma = - 2 i \,\mathrm{n} \cdot \rho, \quad [\mathrm{n}]_{ab,ij} = \delta_{ab,ij} \, n_{ij}
\end{equation}
where $\rho$ is given by the diagonal matrix of phase space volumes \eqref{eq:rhod}, according to \eqref{eq:rhoab}, and $\mathrm{n}$ which we call the ``sheet index'', is a diagonal matrix given by the $n_{ij}$ entries, as denoted above. We represent the multi-sheeted Riemann surface of partial-wave amplitudes in Figure \ref{fig2}.

Note that $N$ distinct square-root thresholds make associated the Riemann surface have $2^N$ sheets in total \cite{Jing:2025qmi}.
Massless exchanges behave differently. If either $m_i = 0$ or $m_j = 0$, additional overlapping branch points arise from the emission of extra massless particles. When both particles are massless, $m_i = m_j = 0$, the branch point collapses to $s = 0$ and may coincide with the left-cut branch point, splitting the physical sheet into two disconnected regions (in Figure \ref{fig}, the red and blue branch points may completely overlap). Any path connecting these regions should then cross extra branch cuts, so the $K$-matrix acquires additional monodromies. In S-matrix bootstrap applications, however, one typically imposes unitarity within restricted sub-sectors of the full multi-particle problem, for example in the jet reduction of \cite{Guerrieri:2024ckc}, so our framework remains meaningful in that context.

Having an explicit analytic continuation of the partial--wave amplitudes to higher sheets, we are now in a position to analyze their analytic structure on those sheets.

\textbf{Higher-sheet analytic structure.} The analytic structure on higher sheets can be described by writing \eqref{eq:Kmatrixsoln} in the form
\begin{equation}
\label{eq:adjdet}
    \mathrm{T}_{(\mathrm{n})}(s) = {\mathrm{adj}\,\big(\mathrm{K}^{-1}(s) - \Sigma_{(\mathrm{n})}(s)\big) \over \det\big(\mathrm{K}^{-1}(s) - \Sigma_{(\mathrm{n})}(s)\big)}
\end{equation}
where $\mathrm{adj}$ denotes the adjugate and $\det$ the determinant. 

The adjugate can be computed using the Cayley–Hamilton theorem \cite{Gantmacher:MatrixTheory1} and is given purely in terms of products of matrix elements of $\mathrm{K}^{-1}(s)$ and $\Sigma_{(\mathrm{n})}(s)$. It can therefore only develop singularities when these individual entries become singular.

 Since $\mathrm{K}^{-1}(s)$ is unchanged by the continuation procedure, it has precisely the same singularities as on the physical sheet, namely those dictated by \eqref{eq:discK}. By contrast, $\Sigma_{(\mathrm{n})}(s)$ given by \eqref{eq:monf} can develop additional singularities upon continuation, such as those in $\rho(s)$ at $s=0$ and $s=(m_i-m_j)^2$, in \eqref{eq:rhod}. These are hidden on the physical sheet, where \eqref{eq:discsigma} only has support for $s \geq (m_i+m_j)^2$, but they can appear on higher sheets. They can be derived independently from a Landau analysis of the bubble integral, where $s=0$ is a second–type singularity and $s=(m_i-m_j)^2$ is the pseudo-threshold singularity \cite{Eden:2002}.

The remaining way for \eqref{eq:Kmatrixsoln} to become singular is when the denominator vanishes at isolated points
\begin{equation}
\label{eq:det}
   \det\big(\mathrm{K}^{-1}(s_0) - \Sigma_{(\mathrm{n})}(s_0)\big)  = 0
\end{equation}
so that $\mathrm{K}^{-1}(s) - \Sigma_{(\mathrm{n})}(s)$ becomes non-invertible as $s \to s_0$. If $s_0 \in \mathbb{R}$, we associate this singularity with a \emph{virtual state}, whereas if $s_0 = s_R \in \mathbb{C}$ has non-zero imaginary part, we associate it with one or more resonances.\footnote{Note that a complex-valued pole on the first sheet would violate the standard analyticity domains expected from causality \cite{Bros:1964fv,Lehmann:1966it}.}

In summary, on higher sheets reached by crossing two-particle cuts, the only possible singularities are the kinematic Landau singularities of the two-particle phase-space and isolated complex poles, which we interpret as resonances or virtual states. No more exotic singularities, such as complex branch points, can appear: there are no ``dragons'' on these unphysical sheets.\footnote{We expect that crossing multi-particle cuts can reveal more intricate structures, such as complex branch points associated with  resonance-particle state production.} We represent the analytic structure of partial-wave amplitudes on higher sheets  in gray in Figure~\ref{fig}.

\textbf{Resonances and their properties.} We now assume the simplest scenario in which the determinant \eqref{eq:det} develops a simple zero at $s = s_R$, to which we associate a resonance,\footnote{A double or higher–order zero would not, in general, imply the factorization properties we derive below, and can be interpreted as two resonance poles colliding. This is analogous to the breakdown of strict factorization when two stable particles become exactly degenerate in mass.}
\begin{equation}
    \det\big(\mathrm{K}^{-1} - \Sigma_{(\mathrm{n})}\big)|_{s \to s_R}  \propto (s - s_R).
\end{equation}
In this case, analyticity implies that as $s \to s_R$ only a single eigenvalue of $\mathrm{K}^{-1} - \Sigma_{{n}}$ vanishes, so that the kernel is one-dimensional. We can then write
\begin{equation}
\label{eq:Kres}
    \big(\mathrm{K}^{-1}(s_R) - \Sigma_{(\mathrm{n})}(s_R)\big)\cdot v =0
\end{equation}
for a unique (up to rescaling) non-trivial vector $v$, whose components can be complex numbers.

Since the adjugate in \eqref{eq:adjdet} is analytic, close to the resonance we  have
\begin{equation}
\label{eq:Tnadj}
      \mathrm{T}_{(\mathrm{n})}(s\to s_R) \propto {\mathrm{adj}\,\big(\mathrm{K}^{-1}(s_R) - \Sigma_{(\mathrm{n})}(s_R)\big) \over s-s_R}\,.
\end{equation}
However, according to \eqref{eq:Kmatrixsoln}, we must have 
\begin{equation}
\label{eq:id}
 \big(\,[\mathrm{K}^{-1}- \Sigma_{(n)}] \cdot \mathrm{T}_{(\mathrm{n})}\,\big)\big|_{s \to s_R} = \mathrm{1}
\end{equation}
and inserting \eqref{eq:Tnadj} requires the pole to cancel out,
\begin{equation}
[\mathrm{K}^{-1}- \Sigma_{(n)}] \cdot \mathrm{adj}\,(\mathrm{K}^{-1}- \Sigma_{(\mathrm{n})}) = 0\,.
\end{equation}
Comparing to \eqref{eq:Kres} it follows that, at $s = s_R$, each column of $ \mathrm{adj}\,(\mathrm{K}^{-1}- \Sigma_{(\mathrm{n})})$ belongs to the kernel of $\mathrm{K}^{-1}- \Sigma_{(\mathrm{n})}$ and is thus proportional to $v$. Given that $v$ is unique and the adjugate is symmetric, we deduce the desired factorization property of the residue in \eqref{eq:Tnadj}
\begin{equation}
    \mathrm{adj}\,\big(\mathrm{K}^{-1}(s_\mathrm{R})- \Sigma_{(\mathrm{n})}(s_\mathrm{R})\big) \propto v \; v^T\,.
\end{equation}
Expanding to sub-leading order $\mathcal{O}((s-s_R)^0)$ in \eqref{eq:id} allows us to fix the proportionality constant (see Appendix \ref{sec:fact} for details) and we find
\begin{equation}
\label{eq:fact}
  \mathrm{T}_{(n)}(s \to s_R) = -{\mathrm{g}_R \; \mathrm{g}_R^T \over s - s_R} + \mathcal{O}((s-s_R)^0)
\end{equation}
with the vector of cubic couplings given by
\begin{equation}
\label{eq:gR}
    \mathrm{g}_R = {v \over \sqrt{v^T \cdot \big({d \over ds}(\Sigma_{(\mathrm{n})} - \mathrm{K}^{-1})|_{s=s_R} \big)\cdot v}}
\end{equation}
where each component corresponds to the coupling of the resonance $R$ to the $ab$ channel, $g_{ab\mathrm{R}} \equiv [\mathrm{g}_\mathrm{R}]_{ab}$.

Equation \eqref{eq:fact} makes manifest the properties of universality and factorization. A resonance, in the sense defined above, generically appears in all channels as a pole of the full S-matrix on an unphysical sheet. Furthermore, the residue factorizes on the pole, providing a definition of the resonance couplings to the various scattering channels.

\textbf{Resonance data and the physical sheet.} Let us now clarify the qualifier ``generically'' used above. A priori there is nothing that forbids some couplings from vanishing, $g_{ijR} = 0$, making the resonance not couple universally to every channel. The key point, however, is that such vanishing couplings impose additional constraints on the amplitude on the physical sheet.

To see this explicitly, we use \eqref{eq:Kmatrixsol} to solve for $\mathrm{K}^{-1}(s)$ in terms of the physical sheet amplitude $\mathrm{T}(s)$, and then reinsert this expression into \eqref{eq:Kres} and \eqref{eq:gR}. In this way we obtain
\begin{equation}
\label{eq:mast1}
    \big(\Sigma_{(\mathrm{n})}(s_R) - \Sigma(s_R) - \mathrm{T}^{-1}(s_R)\big) \cdot v = 0
\end{equation}
with $v$ a non-trivial vector, implying the physical-sheet condition:
\begin{equation}
  \mathrm{det}  \big(\Sigma_{(\mathrm{n})}(s_R) - \Sigma(s_R) - \mathrm{T}^{-1}(s_R)\big) = 0\,.
\end{equation}
The resonance couplings are given by
\begin{equation}
\label{eq:mast2}
    \mathrm{g}_R = {v \over \sqrt{v^T \cdot  \big(\tfrac{d}{ds}(\Sigma_{(\mathrm{n})} - \Sigma - \mathrm{T}^{-1})\big|_{s=s_R} \big)\cdot v}}\,,
\end{equation}
where
\begin{equation}
    \text{Resonance data} = \{\mathrm{n},s_R,\,\mathrm{g}_R\}\,
\end{equation}
with $s_R \in \mathbb{C}$ is the complex pole position of the resonance on sheet $\mathrm{n}$ and $g_{abR} \equiv [\mathrm{g}_R]_{ab} \in \mathbb{C}$ the coupling to the $ab$ two-particle state.

These relations are expressed purely in terms of the partial-wave amplitude on the first sheet, $\mathrm{T}(s)$, and the monodromy $\Sigma_{(\mathrm{n})}(s) - \Sigma(s)$ between the physical sheet and the higher sheet which is given by \eqref{eq:monf}.

In this sense, the properties of universality and factorization follow directly from the physical-sheet data. If a resonance does \emph{not} couple to a given channel, $g_{abR} = 0$, then a component of the eigenvector must vanish, $v_{ab} = 0$,\footnote{Recall that, in our notation, $ab$ corresponds to a single index labeling the $ab$ two-particle state.} which by \eqref{eq:Kres} translates into an additional constraint on the physical-sheet amplitude. Since an analytic function is completely determined by its values on any one of its sheets (for instance via dispersion relations), it is natural that this constraint propagates to all sheets, including the physical one. 

We can express these conditions directly for the partial wave S-matrix $S^J_{ab \to cd}(s)$ defined in \eqref{eq:S}. Namely, in matrix form we have
\begin{equation}
\label{eq:ST}
    \mathrm{S} = 1 + 2 i \,\sqrt{\rho} \cdot \mathrm{T} \cdot \sqrt{\rho}
\end{equation}
where $[\mathrm{S}]_{ab,cd} \equiv S^J_{ab \to cd}$, $\rho$ is the diagonal matrix \eqref{eq:rhoab} and $[1]_{ab,cd} = \delta_{ab,cd}$ is the identity matrix.

We find analogous physical-sheet relations \eqref{eq:mast1} and \eqref{eq:mast2} for the partial-wave S-matrix. Inserting the monodromy \eqref{eq:monf} and the relation \eqref{eq:ST} into \eqref{eq:mast1}  we find the simple formula
\begin{equation}
\label{eq:mastS1}
    \big(\,\mathrm{n} + (\,\mathrm{S}(s_R)-1)^{-1}\,\big) \cdot w = 0
\end{equation}
with $w$ a non-trivial vector, which requires
\begin{equation}
\label{eq:detS}
    \mathrm{det}\big(\mathrm{n} + (\,\mathrm{S}(s_R)-1)^{-1}\big)= 0\,.
\end{equation}
Moreover, if $\tilde{\mathrm{g}}_R \equiv \sqrt{2 i \rho(s_R)} \cdot \mathrm{g}_R$ we find a similarly compact expression for the vector of couplings
\begin{equation}
\label{eq:mastS2}
    \tilde{\mathrm{g}}_R = {w  \over \sqrt{ w^T \cdot \mathrm{n}\cdot \big({d \over ds} \mathrm{S}\big)\big|_{s = s_R} \!\!\cdot \mathrm{n} \cdot w}}\,
\end{equation}
which can take complex values. From the above we deduce the relation
\begin{equation}
\label{eq:gdSg}
    \tilde{\mathrm{g}}_R^T \cdot n \cdot \Big({d \mathrm{S}\over ds} \Big)\bigg|_{s = s_R} \!\!\!\!\!\!\!\!\cdot \mathrm{n} \cdot  \tilde{\mathrm{g}}_R = 1.
\end{equation}
In both results, \eqref{eq:mastS1} and \eqref{eq:mastS2}, all dependence on phase-space factors, including dimension and nature of the branch-cuts drop out, except for the sheet index  $(\mathrm{n})$ specifying the higher-sheet where the resonance is located. These relations can be used directly in S-matrix bootstrap applications (see \cite{He:2023lyy,Tourkine:2023xtu,EliasMiro:2023fqi,Cordova:2023wjp,Guerrieri:2023qbg,Guerrieri:2024jkn,He:2024nwd,Guerrieri:2024ckc,Gumus:2024lmj,Cordova:2025bah,He:2025gws,Correia:2025uvc,deRham:2025vaq} for recent work) and constitute our main practical results. Part of these results, when branch-cuts are of square-root nature, may also be derived via direct analytic continuation of S-matrix unitarity \eqref{eq:uniS} (see Appendix \ref{sec:finapp})

In the next section we illustrate these formulas for a system of two particles with $\mathbb{Z}_2$ symmetry.

\section{example}
\label{sec:examples}
Consider a system of two particles of masses $m_1$ and $m_2$, with $\mathbb{Z}_2$ symmetry where\footnote{We follow the setup of \cite{Homrich:2019cbt} in general space-time dimension.}
\begin{equation}
\label{eq:Tex}
\mathrm{T}(s)=\left(\begin{array}{ll}
T^J_{11\to11}(s) & T^J_{11\to22}(s) \\
T^J_{11\to22}(s) & T^J_{22\to22}(s)
\end{array}\right)
\end{equation}
and the corresponding partial-wave S-matrix elements are given by (\ref{eq:S}). The $\mathbb{Z}_2$ symmetry implies that ${T_{12\to11} = T_{12\to22} = 0}$, and the process $T_{12 \to 12}$ does not couple to either of the channels in \eqref{eq:Tex} in the direct channel, and therefore admits a separate, simpler analysis which we do not consider explicitly here.

\textbf{Pure resonance.} Suppose there is a ``pure'' resonance of spin $J$ at $s = s_R$ in the sheet accessed by encircling the `$11$' cut $n_{11}$ times counterclockwise, without crossing the `$22$' cut, for which $n = \mathrm{diag}(n_{11},0)$. In this case the monodromy \eqref{eq:monf} reduces to
\begin{equation}
\Sigma_{(\mathrm{n})}(s)-\Sigma(s)=-2 i\,\left(\begin{array}{cc}
n_{11} \rho_{11}(s) & 0 \\
0 & 0
\end{array}\right)\,.
\end{equation}
If $d$ is even and $m_1 \neq 0$, then according to \eqref{eq:rhod}, the `$11$' cut will be of square-root nature, and $n_{11} = 0$ (physical sheet) or $n_{11} = 1$ (unphysical sheet). Alternatively, if $d$ is odd or $m_1 = 0$, then $n_{11} \in \mathbb{Z}$ and there are infinitely many higher sheets.

We now analyze the consequences on the physical sheet. Focusing first on the partial-wave amplitudes $\mathrm{T}(s)$, solving the linear system \eqref{eq:mast1} yields the  constraint
\begin{equation}
\label{eq:T11ex}
    T^J_{11\to11}(s_R)= \frac{i}{2\, n_{11}\rho_{11}(s_R)}
\end{equation}
where we take $\mathrm{Im}\, s_R > 0$, thereby fixing the branch of $\rho_{11}(s_R)$. For even $d$ and $m_1\neq 0$, changing the sign of $\mathrm{Im}\, s_R$ flips $\rho_{11}(s_R) \to - \rho_{11}(s_R)$ in \eqref{eq:rhod} due to the square-root branch-cut at $s=4 m_1^2$.
In all other cases, \eqref{eq:T11ex} holds for all $s_R \in \mathbb{C}$.\footnote{Note that for $s \in \mathbb{R}$ the condition \eqref{eq:T11ex} may not be compatible with Hermitian analyticity \eqref{eq:hermitian}, suggesting that the location of virtual states may be constrained.}

The equivalent condition for the partial-wave S-matrix $\mathrm{S}(s)$ follows from \eqref{eq:ST} or \eqref{eq:mastS1} and reads
\begin{equation}
\label{eq:S11ex}
    S^J_{11\to11}(s_R)= 1- \frac{1}{n_{11}}
\end{equation}
where ${n_{11} = 1}$ recovers the usual condition ${ S^J_{11\to11}(s_R) =0}$ derived in \cite{Doroud:2018szp,Guerrieri:2018uew}.

We now compute the physical-sheet relations for the resonance couplings in \eqref{eq:fact}. In the present case $\mathrm{g}_R = (g_{11R}\,,g_{22R})$ and solving \eqref{eq:mast2} gives
\begin{equation}
\label{eq:g11R}
     g_{11R}^2 = - {1 \over \big[2 n_{11}\,\rho_{11}(s_R)\big]^2\,(T_{11\to11}^J)'(s_R)+\,2\, i\, n_{11}\rho_{11}'(s_R) }
\end{equation}
where $'$ denotes a derivative with respect to $s$. Information about $g_{22R}$ is encoded on the physical sheet in the inelastic process $11 \to 22$, which satisfies
\begin{equation}
\label{eq:g22R}
    g_{22R}^2 = - \big[\,2\, g_{11R} \, n_{11} \,\rho_{11}(s_R) \;T^J_{11 \to 22}(s_R) \,\big]^2
\end{equation}
with $g_{11R}$ given by \eqref{eq:g11R}. 

Similarly, for 
\begin{equation}
\tilde{\mathrm{g}}_R = (\tilde{g}_{11R},\tilde{g}_{22R})= (\sqrt{2 i \rho_{11}(s_R)} \, g_{11R}, \sqrt{2 i \rho_{22}(s_R)} \, g_{22R})
\end{equation}
we find simpler relations in terms of the S-matrix on the physical sheet, $\mathrm{S}(s)$. Using e.g. \eqref{eq:mastS2} we find
\begin{equation}
\label{eq:g11tR}
\tilde{g}_{11R}^2 = {1 \over n_{11}^2 \, (S_{11 \to 11}^J)'(s_R)}, \;\;\;\tilde{g}_{22R}^2 = {\big[S^J_{11\to22}(s_R)\big]^2 \over (S^J_{11\to11})'(s_R)}\,.
\end{equation}

Note that setting $\tilde g_{11R} = 0$ would force $(S^J_{11 \to 11})'(s_R)$ to diverge, in conflict with analyticity on the physical sheet. Thus a resonance that crosses only a single cut must necessarily couple to that channel.

By contrast, the relations above show that the `$22$' channel need not couple to the resonance $R$, so $g_{22R} = 0$ is allowed. This imposes a physical-sheet constraint on the inelastic process, $T^J_{11\to22}(s_R)=0$ (equivalently $S^J_{11\to22}(s_R)=0$).

\textbf{Mixed resonance.} We now consider the case where the resonance crosses both the `$11$' and `$22$' cuts, so that
\begin{equation}
\mathrm{n} = \mathrm{diag}(n_{11}, n_{22}) \,, \qquad n_{11} \neq 0 \,,\quad n_{22} \neq 0 \,.
\end{equation}
The corresponding physical–sheet expressions for $\mathrm{T}(s)$ are rather cumbersome, so it is convenient to work instead with the S-matrix $\mathrm{S}(s)$ on the physical sheet.

Using \eqref{eq:detS}, we obtain a symmetric relation involving all channels on the physical sheet,
\begin{align}
\label{eq:mixed1}
\big[&1 - n_{11}\big(1-S^J_{11\to11}(s_R) \big) \big]\!\times\!
\big[1 - n_{22}\big(1-S^J_{22\to22}(s_R)\big)\big] \notag \\
&= n_{11} n_{22} \big[ S^J_{11\to22}(s_R) \big]^2 \,,
\end{align}
which reduces to \eqref{eq:S11ex} when $n_{22} = 0$, and to the analogous relation with $1 \leftrightarrow 2$ when $n_{11} = 0$.

When both cuts are of square-root type then ${n_{11} \in\{0,1\}}$ and ${n_{22} \in\{0,1\}}$. For a resonance that crosses to the second sheet of both channels, ${n_{11} = n_{22} = 1}$, equation \eqref{eq:mixed1} implies the physical-sheet constraint  $S^J_{11\to11}(s_R)\, S^J_{22\to22}(s_R) = \big[ S^J_{11\to22}(s_R) \big]^2$.

Turning now to the couplings of the resonance to the different channels, we solve \eqref{eq:mastS1} for $w$ and substitute into \eqref{eq:mastS2}, obtaining
\begin{equation}
\label{eq:symgs}
\begin{aligned}
    n_{11}\,\tilde{g}^2_{11R}\,\big[1-n_{11}(1-S_{11\to11}^J(s_R))\big]=\\n_{22}\,\tilde{g}^2_{22R}\,\big[1-n_{22}(1-S_{22\to22}^J(s_R))\big]\,.
\end{aligned}
\end{equation}
Equivalently, using \eqref{eq:mixed1}, we can write
\begin{equation}
- \tilde{g}_{11R} \, n_{11} \, S^J_{11\to22}(s_R)  =  \tilde{g}_{22R} \big[1-n_{22}(1-S_{22\to22}^J(s_R))\big]
\end{equation}
which is consistent with \eqref{eq:g11tR} upon setting $n_{22}=0$. An equivalent relation holds  with $1 \leftrightarrow 2$ exchanged, using again  \eqref{eq:mixed1}.

From \eqref{eq:gdSg} we find a final independent relation for the couplings, involving only derivatives of the physical–sheet S-matrix,
\begin{align}
\label{eq:gsderivS}
&n_{11}^2\, \tilde{g}_{11R}^2 \,\bigl(S^{J}_{11\to 11}\bigr)'\!(s_R)
+ n_{22}^2\, \tilde{g}_{22R}^2\, \bigl(S^{J}_{22\to 22}\bigr)'\!(s_R) \notag \\
&\qquad + \;2\, n_{11} n_{22} \,\tilde{g}_{11R}\, \tilde{g}_{22R} \,\bigl(S^{J}_{11\to 22}\bigr)'\!(s_R)
= 1 \,.
\end{align}
Notice that if the resonance does not couple to one of the channels, say $\tilde g_{11R} = 0$, then by \eqref{eq:symgs} it must also satisfy $\tilde g_{22R} = 0$ (and vice versa). In that case, however, relation \eqref{eq:gsderivS} cannot be satisfied, since all matrix elements of $\mathrm{S}'(s_R)$ must remain finite, as required by causality (i.e.\ analyticity on the physical sheet \cite{Lehmann:1958ita}). Consequently, when the resonance crosses both cuts into a higher sheet, $n_{11} \neq 0$ and $n_{22} \neq 0$, it must couple to both channels simultaneously: $g_{11R} \neq 0$ and $g_{22R} \neq 0$.

\section{Discussion}
\label{sec:conc}

In this work we have shown that resonances, defined as complex poles on higher Riemann sheets of the scattering amplitude, share key properties with stable particles. In particular, resonance poles are universal features of the S-matrix, independent of the choice of external scattering states, and they generically appear in every kinematically allowed channel unless excluded by vanishing couplings. Moreover, amplitudes exhibit factorization across resonance poles (see equation \eqref{eq:fact}), in close analogy with the well-known factorization at stable particle poles.  

Within clearly stated assumptions on analyticity and unitarity, our analysis provides a unified framework for resonance poles on higher Riemann sheets, reproducing the familiar second-sheet picture \cite{Gribov:2009cfk,Guo:2015daa} and extending it to arbitrary higher sheets reachable by two-particle cuts and to arbitrary spacetime dimensions, including the case of odd dimensions and massless exchanges where the kinematic Riemann surface is infinitely sheeted.

Crucially, we also demonstrated that resonance data, including the mass, width, and couplings of the resonance, is in principle fully encoded in the first (physical) sheet, even when the pole itself lies on a more distant higher sheet (see equations \eqref{eq:mast1} and \eqref{eq:mast2}). This observation is especially relevant for S-matrix bootstrap applications, where only first-sheet information is directly accessible. We expect the results derived here to be useful for future coupled-channel and multi-particle S-matrix bootstrap studies (see \cite{Homrich:2019cbt,Guerrieri:2024ckc} for work in this direction).

In the example studied in Sec.~\ref{sec:examples}, we further saw that causality, encoded as analyticity in the physical sheet, can impose additional constraints on the resonance data. In particular, a resonance that does not couple to a given channel cannot cross the corresponding branch cut into a higher Riemann sheet.\footnote{The converse statement is not true: A resonance can couple to a channel whose branch-cut it did not cross.} 

In this work we focused on generic $2\to2$ amplitudes with external states $a,b \to c,d$ and considered resonances in higher sheets of two-particle thresholds. The properties established above apply to any resonance on any sheet that can be reached through a sequence of two-particle cuts. Although multi-particle cuts were not treated explicitly, our results already accommodate an arbitrary number of intermediate two-particle states. In light of \cite{Guerrieri:2024ckc}, which shows that multi-particle states can be reorganized into an infinite collection of effective two-particle states labeled by an additional ``jet'' quantum number, it is possible that our conclusions generalize to that setting as well.

An important consequence of our analysis is that, on higher sheets of partial-wave amplitudes connected only by two-particle cuts, no singularities other than isolated poles and the Landau singularities associated with two-body phase-space can occur (see Figure \ref{fig}). In other words, on these sheets there are no additional exotic singularities beyond those dictated by unitarity. In this sense, there are no ``dragons'' lurking on these higher sheets.\footnote{Momentum-space scattering amplitudes are expected to have additional Landau singularities on higher sheets, associated with the non-convergence of the partial-wave decomposition \cite{Lehmann:1958ita,Correia:2020xtr}. In particular, natural boundaries are expected to occur \cite{Freund:1961NaturalBoundary,Mizera:2022dko}.}

These results suggest several natural directions for future investigation.

\textbf{Complex production thresholds.}
On Riemann sheets connected to multi-particle cuts involving three or more particles, the analytic structure becomes substantially richer. One naturally expects isolated poles associated with resonances that couple directly to such multi-particle states, for instance the $\eta$ meson, which predominantly decays into three neutral pions~\cite{ParticleDataGroup:2024cfk}. In addition, complex-valued branch points arise at production thresholds of multi-particle states containing at least one unstable constituent \cite{Ceci:2011ae}. It would be interesting to understand how these complex branch points are represented after the two-particle jet reduction of \cite{Guerrieri:2024ckc}. 

By contrast, extending the simple K-matrix framework used here across two-particle cuts to a formulation that also remains valid in the presence of genuine multi-particle cuts is far from straightforward; see \cite{Hansen:2014elt,Romero-Lopez:2019qrt,Hansen:2021ofl} for progress in this direction. On the other hand, the relations \eqref{eq:mastS1} and \eqref{eq:mastS2} are completely independent of phase-space factors and of the spacetime dimension, suggesting that they may admit a direct generalization to multi-particle channels. In perturbation theory such channels typically involve elliptic integrals  \cite{Zamolodchikov:2011wd,Broedel:2019kmn} and are therefore expected to give rise to infinitely sheeted Riemann surfaces in any spacetime dimension, a structure that our framework already accommodates within the two-particle sector.

\textbf{Higher-point factorization.}
The results in this paper were derived for amplitudes with four external particles. A natural extension is to analyze higher-point amplitudes and to show that resonance poles exhibit factorization not only in their cubic couplings but also in higher-point sub-amplitudes. Early work in this direction can be found in \cite{Gross:1965NormalThreshold,Iagolnitzer:1977PoleFactorization}.

\textbf{Bounds on resonance couplings.}
Given the expressions derived here that relate resonance data to first-sheet information, it would be natural to use the S-matrix bootstrap to bound resonance couplings. In particular, it would be interesting to investigate whether, in contrast to couplings of stable particles, resonance couplings can take opposite values and generate higher-order poles or even cancel each other. See  \cite{Doroud:2018szp} for related work in the context of two space-time dimensions.

From a phenomenological point of view, resonance couplings are often poorly determined experimentally, which provides further motivation for a bootstrap approach. For example, the $g_{\sigma \pi \pi}$ coupling of the broad $\sigma$ meson, which decays into two pions, does not yet appear to have been measured \cite{ParticleDataGroup:2024cfk}.

\textbf{Experimental identification of resonances.}
Identifying resonances from experimental data is intrinsically challenging, since measurements are restricted to the real physical region. For narrow resonances close to the real axis, the nearby pole cleanly dominates the amplitude and a simple Breit–Wigner parameterization is often adequate~\cite{ParticleDataGroup:2024cfk}. For broad resonances, or in the presence of overlapping structures and nearby thresholds, the situation is far more complicated. In practice, one typically performs multi-channel amplitude analyses using the K-matrix or related dispersive parameterizations, but these fits are highly nontrivial and model dependent~\cite{JPAC:2021rxu,Mai:2022eur}. 
A fully rigorous extraction of resonance parameters, with systematic uncertainties under quantitative control, remains a major open problem in hadron spectroscopy~\cite{JPAC:2022ipt}.

\textbf{Resonances in the S-matrix bootstrap.}
In this context, modern S-matrix bootstrap approaches~\cite{Paulos:2016but,Paulos:2017fhb,Tourkine:2023xtu,Gumus:2024lmj} stand out, to our knowledge, as the only explicit constructions of scattering amplitudes that manifestly satisfy unitarity, analyticity, and crossing symmetry at all energies. The emergence of resonances and Regge trajectories in the resulting amplitudes is by now well established~\cite{Guerrieri:2018uew,Haring:2022sdp,Acanfora:2023axz,Guerrieri:2023qbg,Cordova:2023wjp,He:2023lyy,Gumus:2024lmj,Guerrieri:2024jkn,He:2024nwd,Correia:2025uvc,He:2025gws}, and it is natural to expect that, in the foreseeable future, S-matrix bootstrap techniques will evolve into a standard tool for extracting resonance properties directly from experimental or lattice inputs, as has already been initiated in \cite{Guerrieri:2024jkn}. The physical-sheet relations for resonance data derived in this work are well suited for such applications and should prove useful in phenomenological resonance identification based on the S-matrix bootstrap.

\vspace{2mm}

\emph{Acknowledgments.}  We are particularly grateful to Aditya Hebbar for discussions and collaboration on related work. We thank Simon Caron-Huot, Sérgio Carrôlo, Alessandro Georgoudis, Andrea Guerrieri, Kelian Häring, Alexandre Homrich, Giulia Isabella, Kyle Lee, Ian Moult, Julia Pasiecznik, João Penedones, David Poland, Balt van Rees, Pedro Vieira, and Sasha Zhiboedov for useful discussions and comments on the draft. The authors are supported by the
National Science and Engineering Council of Canada (NSERC) and the Canada Research Chair program, reference number CRC-2022-00421. C.P.~is additionally supported by the Walter C.~Sumner Memorial Fellowship.

\begin{appendix}

\section{Kinematics and two-particle unitarity}
\label{sec:kin}

\textbf{Center-of-mass kinematics.}
We consider a generic ${2 \rightarrow 2}$ scattering process with Mandelstam invariants
\begin{equation}
\begin{aligned}
s=\left(p_a+p_b\right)^2, \quad &t=\left(p_a-p_c\right)^2, \quad u=\left(p_a-p_d\right)^2\\
s+t+u&=m_a^2+m_b^2+m_c^2+m_d^2\,.
\end{aligned}
\end{equation}
We take $\theta$ to be the physical scattering angle between incoming particle $a$ and outgoing particle $c$ and define
$z=\cos \theta=\frac{\vec{p}_a \cdot \vec{p}_c}{\left|\vec{p}_a\right|\left|\vec{p}_c\right|}$. 
Energy–momentum conservation fixes the energies and $(d-1)$–momenta in terms of $s$ and the scattered masses
\begin{equation}
\begin{aligned}
&\begin{array}{ll}
E_a=\frac{s+m_a^2-m_b^2}{2 \sqrt{s}}, & E_b=\frac{s+m_b^2-m_a^2}{2 \sqrt{s}} \\
E_c=\frac{s+m_c^2-m_d^2}{2 \sqrt{s}}, & E_d=\frac{s+m_d^2-m_c^2}{2 \sqrt{s}}
\end{array}
\end{aligned}
\end{equation}
\begin{equation}
    \begin{aligned}
    &\left|\vec{p}_a\right|=\left|\vec{p}_b\right|=\frac{\sqrt{\lambda\left(s, m_a^2, m_b^2\right)}}{2 \sqrt{s}},\\
    &\left|\vec{p}_c\right|=\left|\vec p_d\right|=\frac{\sqrt{\lambda\left(s, m_c^2, m_d^2\right)}}{2 \sqrt{s}}
    \end{aligned}
\end{equation}
where the the Källén function is
\begin{equation}
    \lambda(x,y,z)
= x^2 + y^2 + z^2 - 2(xy + yz + zx)\,.
\end{equation}
We have 
\begin{equation}
t=m_a^2+m_c^2-2 E_a E_c+2\left|\vec{p}_a\right|\left|\vec{p}_c\right| \cos \theta\,.
\end{equation}
Writing the angle in terms of masses and Mandelstam invariants we obtain
\begin{equation}
z = 
\frac{
2s\bigl(t - m_a^2 - m_c^2\bigr)
+ \bigl(s + m_a^2 - m_b^2\bigr)\bigl(s + m_c^2 - m_d^2\bigr)
}{
\sqrt{\lambda\bigl(s,m_a^2,m_b^2\bigr)\,\lambda\bigl(s,m_c^2,m_d^2\bigr)}
} .
\end{equation}

\textbf{Two-particle unitarity.}
The Lorentz-invariant two-particle on-shell phase-space measure, represented in \eqref{eq:uniT}, is given explicitly by
\begin{equation}
    \int_{ij} \equiv b_{ij} \int\frac{d^{d-1} \vec{q_i}}{(2\pi)^{d-1}\,2 E_{i}} \frac{d^{d-1} \vec{q_j}}{(2\pi)^{d-1} 2 E_{j}}
\end{equation}
where $b_{ij} = 1-\delta_{ij}/2$ is the Bose symmetry factor, and $E_{i}=\sqrt{q_i^2+m_i^2}$ and $E_{j}=\sqrt{q_j^2+m_j^2}$ are the on-shell energies of the intermediate particles of masses $m_i$ and $m_j$, respectively.
Considering only a single two-particle contribution to \eqref{eq:uniT} we have
\begin{equation}
\begin{aligned}
2 \operatorname{Im} T_{a b \rightarrow c d} \supset & \,b_{ij} \int\frac{d^{d-1} \vec{q_i}}{(2\pi)^{d-1}\,2 E_{i}} \frac{d^{d-1} \vec{q_j}}{(2\pi)^{d-1} 2 E_{j}}(2 \pi)^{d} \\
&\times\delta^d\left(p_a+p_b-q_i-q_j\right) T_{a b \rightarrow i j}^* T_{i j \rightarrow c d}\,.
\end{aligned}
\end{equation}
In the center-of-mass frame, where $p_a^\mu+p_b^\mu=(\sqrt{s}, \vec{0})$, we factor the momentum conserving delta function 
\begin{equation}
\delta^d\left(p_a+p_b-q_i-q_j\right)=\delta\left(\sqrt{s}-E_{i}-E_{j}\right) \delta^{(d-1)}\left(\vec{q}_i+\vec{q}_j\right).
\end{equation}
The spatial delta-function enforces $\vec{q}_i=-\vec{q}_j$ and we define $\vec{q}_i\equiv |\vec{q}| \vec{n}$ with $\vec{n}$ a unit vector in $(d-1)$.
Introducing spherical coordinates for the remaining momentum integral $d^{d-1} \vec{q}_i=|\vec{q}|^{d-2} d |\vec{q}| d^{d-2} \Omega_{\vec{n}}$, we obtain
\begin{equation}
\label{eq:2Tdeltap}
\begin{aligned}
2 \operatorname{Im} T_{a b \rightarrow c d}(s,t)= b_{ij}(2 \pi)^{2-d}&\int_0^{\infty} d|\vec{q}| |\vec{q}|^{d-2} \int d^{d-2} \Omega_{\vec{n}} \\
&\times\frac{\delta\left(|\vec{p}_i|-|\vec{q}|\right)}{4 |\vec{q}| \sqrt{s}}T_{a b \rightarrow i j}^* T_{i j \rightarrow c d}\,
\end{aligned}
\end{equation}
where $|\vec{p}_i|^2=\frac{\lambda\left(s, m_i^2, m_j^2\right)}{4 s}$. Thus, we find
\begin{equation}
\label{eq:uniTT}
\begin{aligned}
    2 \operatorname{Im} T_{a b \rightarrow c d}(s,t)=&\frac{b_{ij}}{4 (4\pi)^{d-2}}s^{\frac{2-d}{2}}\lambda\left(s, m_i^2, m_j^2\right)^{\frac{d-3}{2}}\\
    &\times\int d^{d-2} \Omega_{\vec{n}}T_{a b \rightarrow i j}^* T_{i j \rightarrow c d}\,.
\end{aligned}
\end{equation}
The final angular integral may be written in terms of the external and internal scattering angles as
\begin{equation}
\int d^{d-2} \Omega_{\vec{n}} \equiv \int_{-1}^1 d z^{\prime} \int_{-1}^1 d z^{\prime \prime} \mathcal{P}_d\left(z, z^{\prime}, z^{\prime \prime}\right)
\end{equation}
where the kernel $\mathcal{P}_d$ obeys \cite{Correia:2020xtr}
\begin{equation}
\label{eq:kernel}
\begin{aligned}
\mathcal{P}_d\left(z, z^{\prime}, z^{\prime \prime}\right)=&(4 \pi)^{d-2} \mathcal{N}_d^2\left(1-z^{\prime 2}\right)^{\frac{d-4}{2}}\left(1-z^{\prime \prime 2}\right)^{\frac{d-4}{2}} \\
&\times\sum_{L=0}^{\infty} n_L^{(d)} P_L^{(d)}(z) P_L^{(d)}\left(z^{\prime}\right) P_L^{(d)}\left(z^{\prime \prime}\right)\,.
\end{aligned}
\end{equation}
Projecting \eqref{eq:uniTT} onto fixed angular momentum using the partial-wave decomposition \eqref{eq:Tpwd}, we obtain 
\begin{equation}
\label{eq:uniTT2}
\begin{aligned}
\operatorname{Im} &T^J_{a b \rightarrow c d}(s)
= b_{ij}\frac{s^{\frac{2-d}{2}}\lambda\!\left(s,m_i^2,m_j^2\right)^{\frac{d-3}{2}}}{4(4\pi)^{d-2}}
\frac{\mathcal{N}_d}{2}\\
&\times\int dz' dz''\,
T_{a b \rightarrow i j}^*(s,t(z')) T_{i j \rightarrow c d}(s,t(z''))
\\
&\qquad\times
\int dz\,(1-z^2)^{\frac{d-4}{2}}
P_J^{(d)}(z)\,
\mathcal{P}_d\!\left(z,z',z''\right).
\end{aligned}
\end{equation}
We combine \eqref{eq:kernel} with the orthogonality relation 
\begin{equation}
\frac{1}{2} \int_{-1}^1 d z\left(1-z^2\right)^{\frac{d-4}{2}} P_J^{(d)}(z) P_L^{(d)}(z)=\frac{\delta_{J L}}{\mathcal{N}_d n_J^{(d)}}
\end{equation}
and employ the partial wave expansions of $T_{a b \rightarrow i j}^*$ and $ T_{i j \rightarrow c d}$ in \eqref{eq:uniTT}. This yields
\begin{equation}
\begin{aligned}
    \operatorname{Im} T_{a b \rightarrow c d}=&b_{ij} s^{\frac{2-d}{2}}\left(s-\left(m_i-m_j\right)^2\right)^{\frac{d-3}{2}}\\
    &\times\left(s-\left(m_i+m_j\right)^2\right)^{\frac{d-3}{2}}  \!\left(T_{a b \rightarrow i j}^J\right)^* T_{i j \rightarrow c d}^J\,
    \end{aligned}
\end{equation}
reproducing (\ref{eq:uniP}) and (\ref{eq:rhod}).

\section{Relation to CDD solutions, form factors and Watson's theorem}
\label{sec:cdd}

\textbf{CDD and K-matrix solutions.} The  CDD \cite{Castillejo:1955ed} solution of elastic unitarity, say for a process $a b \to a b$, takes the form
\begin{equation}
\label{eq:CDD}
    S^\text{CDD}_{ab \to ab}(s) = \prod\limits_{i} \frac{c_i+i\rho_{ab}(s)}{c_i-i\rho_{ab}(s)},\quad  s \geq (m_a+m_b)^2,
\end{equation}
where $c_i \in \mathbb{R}$. It is clear that the CDD solution will satisfy elastic unitarity $|S^\text{CDD}_{ab \to ab}(s)|^2 = 1$. Moreover it can be analytically continued into the complex plane using the explicit expression for the phase-space factor \eqref{eq:rhod}.

Now, squaring the phase-space volume \eqref{eq:rhod} we see that all square-root branch-cuts cancel and $[\rho_{ab}(s)]^2$ is a rational function of $s$. It thus becomes clear that the product in the CDD solution \eqref{eq:CDD} can always be written in the K-matrix form 
\begin{equation}
    S^\text{CDD}_{ab \to ab}(s) =  \frac{K^{-1}(s)+i\rho_{ab}(s)}{K^{-1}(s)- i \rho_{ab}(s)},
\end{equation}
with $K^{-1}(s)$ a rational function and, in particular, without any branch-cut across $s \geq (m_i+m_j)^2$. See also Appendix A of \cite{Correia:2022dyp}.

\textbf{Form factors and Watson's theorem.} Given a local operator  $\mathcal{O}(x)$ the associated two-particle form factor $F_\mathcal{ab}(s)$ satisfies the elastic unitarity relation \cite{Karateev:2019ymz,He:2023lyy}
\begin{equation}
\mathrm{Im}\, F_{ab}(s) = \rho_{ab}(s) [T_{ab \to ab}(s)]^*  F_{ab}(s).
\end{equation}
Comparing with \eqref{eq:uniT} we see that the form factor obeys the same unitarity relation as the amplitude for a mixed process $T_{ab \to cd}$ with $\{c,d\} \neq \{a,b\}$, across the $\{a,b\}$ cut. Form factors thus admit a general K-matrix solution, and our general results derived for resonances in S-matrix elements should also apply for form factors. 

In particular, letting
\begin{equation}
\label{eq:Texapp}
\tilde{\mathrm{T}}(s)\equiv\left(\begin{array}{ll}
T_{ab\to ab}(s) & F_{ab}(s) \\
F_{ab}(s) & \Pi(s)
\end{array}\right)
\end{equation}
where $\Pi(s)$ is the time-ordered two-point function of local operator $\mathcal{O}(x)$ in momentum space. We find that unitarity for this object takes the same form as \eqref{eq:uniF2}
\begin{equation}
\label{eq:uniF2app}
    \mathrm{Disc}\, \tilde{\mathrm{T}}
    = (\tilde{\mathrm{T}})^{\dagger} \!\cdot \tilde{\rho}  \cdot \!\tilde{\mathrm{T}}
\end{equation}
with $\tilde{\rho} = \mathrm{diag}(\rho_{ab},0)$.

We then have
\begin{equation}
    \tilde{\mathrm{T}}(s)  = [\tilde{\mathrm{K}}^{-1}(s) - \tilde{\Sigma}(s)]^{-1}.
\end{equation}
with $\tilde{\Sigma}(s) = \mathrm{diag}(\Sigma_{ab}(s), 0)$ respecting \eqref{eq:discsigma}, and $\mathrm{Disc} \,\tilde{\mathrm{K}}(s) = 0$. Taking 
\begin{equation}
\label{eq:Kex}
\tilde{\mathrm{K}}(s)\equiv\left(\begin{array}{ll}
\alpha(s) & \beta(s) \\
\beta(s) & \sigma(s)
\end{array}\right)
\end{equation}
we find
\begin{equation}
    T_{ab \to ab}(s) = {\alpha(s) \over 1 - \alpha(s) \,\Sigma_{ab}(s)},
\end{equation}
and
\begin{equation}
F_{ab}(s) = {\beta(s) \over 1 - \alpha(s) \,\Sigma_{ab}(s)}.
\end{equation}
Now, for $s \geq (m_a + m_b)^2$ the function $\Sigma_{ab}(s)$ acquires an imaginary part while $\alpha(s)$ and $\beta(s)$ remain real. This implies Watson's theorem \cite{Watson:1952ji}:
\begin{align}
{F_{ab}(s) \over F^*_{ab}(s)} &= {1 - \alpha(s) \,\Sigma_{ab}^*(s) \over 1 - \alpha(s) \,\Sigma_{ab}(s)} \notag \\ 
&= 1+ 2 i \rho_{ab}(s) T_{ab \to ab}(s) = S_{ab\to ab}(s).
\end{align}

\section{Derivation of factorized residue on resonance pole}
\label{sec:fact}
Here we derive the factorization property \eqref{eq:fact} and the proportionality constant in relation \eqref{eq:gR}. 

If $A(s)$ is the inverse of $T(s)$ we have
\begin{equation}
\label{eq:TA}
    T(s) \cdot A(s) = 1
\end{equation}
where 
\begin{equation}
\label{eq:Av}
    A(s_R) \cdot v = A_0 \cdot v = 0,
\end{equation}
for a non-trivial vector $v$. Around the resonance pole we have
\begin{equation}
    T(s \sim s_R) = {T_{-1} \over s- s_R} + T_0 + \dots
\end{equation}
and
\begin{equation}
    A(s \sim s_R) = A_{0} + A_1 (s - s_R) + \dots
\end{equation}
Plugging these expansions into \eqref{eq:TA} we find
\begin{equation}
\label{eq:AT2}
   A_0  \, T_{-1} = 0, \qquad  A_1 \, T_{-1}  +  A_0 \,T_0 = 1.
\end{equation}
From \eqref{eq:Av} and the first equation of \eqref{eq:AT2}  we find that the residue must factorize in terms of $v$, where symmetry $T_{-1}^T = T_{-1}$ implies the form
\begin{equation}
\label{eq:vvT}
 T_{-1} = \alpha\, v\, v^T
\end{equation}
with $\alpha$ a proportionality constant we will fix now.  We take the second equation of \eqref{eq:AT2} and multiply on the left by the vector $v^T$ and use $v^T A_0  = 0$ to write
\begin{equation}
    v^T A_1 T_{-1} = v^T
\end{equation}
We now plug \eqref{eq:vvT} into the above and get 
\begin{equation}
   \alpha \,v^T A_1 v \,v^T = v^T
\end{equation}
which implies
\begin{equation}
    \alpha = {1 \over v^T A_1 v}
\end{equation}
and inserting into \eqref{eq:vvT} we recover the equation \eqref{eq:gR} of the main text.

\section{Physical-sheet conditions from analytic continuation of unitarity}
\label{sec:finapp}

Here we showcase how analytic continuation of the unitarity relation \eqref{eq:uniS} for the S-matrix can recover part of the results derived in Section \ref{sec:anacont}. Making use \eqref{eq:ST}, we have that unitarity \eqref{eq:uniS} reads in matrix form, with explicit $i \epsilon$ prescription,
\begin{equation}
\label{eq:S2}
    \mathrm{S}^\dagger(s+ i \epsilon) \cdot\mathrm{S}(s+i\epsilon) = 1 - \Delta'(s)
\end{equation}
where $0\preceq\Delta' \preceq 1$, so that $\mathrm{S}^\dagger \mathrm{S} \preceq 1$. 

We proceed to evaluate \eqref{eq:S2} where $\Delta'$ has no support, in the two-particle region (depicted in blue in Figure \ref{fig}). Moreover, if the two-particle cuts are of square-root type, we find that real analyticity of the amplitude $\mathrm{T}(s)$ in \eqref{eq:hermitian} translates into the partial-wave S-matrix $\mathrm{S}(s)$ via \eqref{eq:ST} since $\big[i \rho(s)\big]^* = - i \rho(s^*) = i \rho(s)$. 

Using $\mathrm{S}^*(s) = \mathrm{S}(s^*)$ in \eqref{eq:S2} we find the manifestly analytic relation 
\begin{equation}
\label{eq:S3}
    \mathrm{S}^T(s- i \epsilon) \cdot\mathrm{S}(s+i\epsilon) = 1
\end{equation}
which we can analytically continue to complex $s$, and we find the familiar relation from \cite{Doroud:2018szp,Guerrieri:2018uew} in matrix form,
\begin{equation}
\label{eq:uniana}
    \mathrm{S}_{(1)}^T(s) \cdot\mathrm{S}(s) = 1
\end{equation}
where $\mathrm{S}_{(1)}^T(s)$ is evaluated in the higher-sheet given by crossing all the two-particle cuts, $\mathrm{n} = 1$, with $1$ the identity matrix.

Assuming a resonance pole 
\begin{equation}
    \mathrm{S}_{(1)}^T(s\sim s_R) = {\mathrm{\tilde{g}}_R \,\mathrm{\tilde{g}}_R^T  \over s - s_R}
\end{equation}
and inserting into \eqref{eq:uniana} requires, at leading order,
\begin{equation}
\label{eq:szero}
    \mathrm{S}(s_R) \cdot \mathrm{\tilde{g}}_R = 0.
\end{equation}
with non-trivial vector of couplings $ \mathrm{\tilde{g}}_R$,
and we must have $\det \mathrm{S}(s_R) = 0$. The above relation is recovered from \eqref{eq:mastS1} by taking $\mathrm{n} = 1$ and multiplying on the left-hand-side by $\mathrm{S}(s_R) - 1$. Likewise, expanding \eqref{eq:uniana} to sub-leading order we recover relation \eqref{eq:gdSg} with $\mathrm{n} = 1$. 

Obtaining relations \eqref{eq:mastS1} and \eqref{eq:mastS2} in full generality with $\mathrm{n} = \mathrm{diag}(n_{ij})$, where $n_{ij} \in \mathbb{Z}$, is unclear from direct continuation of unitarity \eqref{eq:uniS}, since $\mathrm{S}(s)$ is not necessarily real analytic for logarithmic type branch-points. 

However, even in this case, a relation between pole on the nearest unphysical sheet $\mathrm{n} = 1$ and a ``zero'' on the physical sheet $\mathrm{n} = 0$, \eqref{eq:uniana}, can still be derived from unitarity $\mathrm{S}^\dagger(s) \mathrm{S}(s) = 1$ via the replacement, valid for real energies, $[\mathrm{S}(s)]^\dagger = [\mathrm{S}(s^*)]^\dagger$. Given that $[\mathrm{S}(s^*)]^\dagger$ is now an analytic function of $s$, the unitarity relation $[\mathrm{S}(s^*)]^\dagger \mathrm{S}(s) = 1$ may be analytically continued into the complex plane where, taking $\mathrm{Im} \,s > 0$, we see that $s^*$ goes into the higher-sheet, with index $\mathrm{n}= -1$, and we find the relation
\begin{equation}
    [\mathrm{S}^{(-1)}(s^*)] \cdot \mathrm{S}(s) = 1
\end{equation}
meaning that a pole for $s = s_R$ in the sheet with index $\mathrm{n} = -1$ implies a zero on the physical-sheet for $s = (s_R)^*$: $\mathrm{S}(s_R^*) \cdot \tilde{\mathrm{g}}_R = 0$.

If the branch-cut is of square-root type then $\mathrm{S}(s)$ would be real analytic which would imply also the usual zero for $s = s_R$ derived in \eqref{eq:szero}. On the other hand, if the branch-cut is of logarithmic type, we would have a different condition for $s = s_R$ given by solving \eqref{eq:mastS1} with $\mathrm{n} = -1$. In the simplest case of crossing a single branch-cut (see example in Sec. \ref{sec:examples}), we apply condition \eqref{eq:S11ex} with $n_{11} = -1$ giving $S_{11 \to 11}(s_R) = 2$ on the physical-sheet.

\end{appendix}

\bibliographystyle{JHEP}
\bibliography{refs}  

\end{document}